\documentclass[onecollarge,natbib]{svjour2}
\bibpunct{[}{]}{;}{n}{}{,} 
\smartqed  
\usepackage{graphicx}
\usepackage{braket}
\usepackage{amsmath}

\newcommand{\inst}{\hbox{\scriptsize inst}}
\newcommand{\eff}{\text{eff}}

\newcommand{\tot}{\text{tot}}

\journalname{Few-Body Systems}
\begin{document}

\title{Non-perturbative Calculation of the Positronium Mass Spectrum in Basis Light-Front Quantization
\thanks{Presented at LightCone 2014, Raleigh, North Carolina.}}

\titlerunning{Positronium Spectrum in BLFQ}        

\author{Paul Wiecki \and
        Yang Li \and
        Xingbo Zhao \and
        Pieter Maris        \and
        James P. Vary
}

\authorrunning{Wiecki \it{et al}} 

\institute{P.~Wiecki \and Y.~Li \and X.~Zhao \and P.~Maris \and J.~P.~Vary \at
              Department of Physics and Astronomy, Iowa State University, Ames, IA 50011, USA \\
              \email{pwwiecki@iastate.edu}
}

\date{\today}

\maketitle

\begin{abstract}
We report on recent improvements to our non-perturbative calculation of the positronium spectrum.
Our Hamiltonian is a two-body effective interaction which incorporates one-photon exchange terms,
but neglects fermion self-energy effects. This effective Hamiltonian is diagonalized numerically in a 
harmonic oscillator basis at strong coupling ($\alpha=0.3$) to obtain the mass eigenvalues. 
We find that the mass spectrum compares favorably to the Bohr spectrum of non-relativistic quantum
mechanics evaluated at this unphysical coupling.
\end{abstract}

\section{Introduction}
\label{intro}
Basis Light-Front Quantization (BLFQ) \cite{vary} is a non-perturbative tool for solving bound state problems in quantum field theory.
This Hamiltonian-based approach  
combines the advantages of light-front dynamics \cite{brodskyreview,harinotes}
with modern developments in {\it ab initio} nuclear structure calculations, such as the
No-Core Shell Model (NCSM) \cite{ncsm,Navratil:2000gs}. 
In BLFQ, the quantum field theoretical bound state problem is formulated as a large, sparse matrix diagonalization problem. State-of-the-art methods developed for
NCSM calculations can then be used to address hadronic systems \cite{methods1,methods2,methods3}.
The diagonalization of the light-front Hamiltonian in a Fock-space basis yields the mass eigenstates of the system, along with amplitudes for evaluating 
non-perturbative observables.
Here, we extend previous work \cite{wiecki} on the mass spectrum of positronium in BLFQ. We evaluate
the spectrum at a strong coupling of $\alpha=0.3$ to make both
the binding energy and hyperfine splitting easier to resolve numerically.
Details of the calculations and additional results can be found in Ref.  \cite{forthcoming}.

\section{Basis Light-Front Quantization}
\label{sec:blfq}
In BLFQ, hadron observables are evaluated by solving the eigenvalue equation
$
P^\mu P_\mu \ket{\Psi}=M^2 \ket{\Psi},
\label{eq:bound_st}
$
where $P^\mu$ is the energy-momentum 4-vector operator and $M$ is the invariant mass. In BLFQ, we express the operator $P^2$ in 
light-cone gauge. The operator $P^2$ then plays a role analogous to the 
Hamiltonian operator in non-relativistic quantum mechanics. As such, it is sometimes referred to as the
``light-cone Hamiltonian'' $H_{LC}\equiv P^2$. (Note that in this convention the ``Hamiltonian'' has energy squared
units.) This operator can be derived from any field theoretical Lagrangian via the Legendre transform. 
In BLFQ, Eq.~(\ref{eq:bound_st}) is expressed in a truncated basis, and the resulting finite-dimensional matrix is diagonalized numerically.
One then examines the trends in observables as the basis truncation is relaxed to estimate the results in the 
infinite matrix (or ``continuum'') limit.
In BLFQ, three separate basis truncations are made.

First, the number of Fock sectors in the basis is truncated. 
The basis space for the diagonalization in principle includes an infinite set of Fock sectors.
For example,
the positronium wavefunction could be expressed schematically as
$
\Ket{e^+e^-}_{\hbox{\scriptsize phys}}=a\Ket{e^+e^-}+b\Ket{e^+e^-\gamma}+c\Ket{e^+e^-\gamma\gamma}+d\Ket{\gamma}+f\Ket{e^+e^-e^+e^-}+\cdots.
$
Here we limit ourselves
to only the $\Ket{e^+e^-}$ and $\Ket{e^+e^-\gamma}$ sectors. We do not yet make any attempt to examine the limit of increasing the 
number of Fock sectors.

Secondly, we discretize the longitudinal momentum by putting our
system in a longitudinal box of length $L$ and applying (anti-)periodic boundary conditions (BCs). Specifically, we choose
periodic BCs for bosons and anti-periodic BCs for fermions. Thus
$
p^+=\frac{2\pi}{L}j,
$
where $j$ is an integer for bosons, or a half-integer for fermions. For bosons, we exclude the ``zero modes'', i.e. $j\neq0$. 
In the many-body basis, all basis states are selected to have the same total longitudinal momentum $P^+=\sum_ip_i^+$,
where the sum is over the particles in a particular basis state. We then parameterize $P^+$ using a dimensionless variable $K=\sum_i j_i$ such that 
$P^+=\frac{2\pi}{L}K$. For a given particle $i$, the longitudinal momentum fraction $x$ is defined as
$
x_i=p_i^+/P^+=j_i/K.
$
Due to the positivity of longitudinal momenta on the light-front \cite{harinotes}, fixing $K$ also serves as
a Fock space cutoff and makes the number of longitudinal modes finite \cite{1+1}. It is easy to see that $K$
determines our ``resolution'' in the longitudinal direction, and thus our resolution on parton distribution functions.
The longitudinal continuum limit corresponds to the limit $L,K \to \infty$.

Finally, in the transverse direction, we employ a 2D Harmonic Oscillator (HO) basis. 
The basis functions are the eigenfunctions of the operator ($\mathbf{q}_i\equiv\mathbf{p}_i/\sqrt{x_i}$ and $\mathbf{s}_i\equiv\sqrt{x_i}\mathbf{r}_i$)
\begin{equation*} 
P_+^\Omega=
\sum_i\left(\frac{\mathbf{p}_i^2}{2p_i^+}+\frac{\Omega^2}{2}p_i^+\mathbf{r}_i^2   \right)=
\frac{\Omega}{2}\sum_i\left[\left(\frac{\mathbf{q}_i}{\sqrt{P^+\Omega}}\right)^2+\left(\sqrt{P^+\Omega}\,\mathbf{s}_i\right)^2   \right]
\end{equation*}
The eigenfunctions  $\Psi_{n_i}^{m_i}(\mathbf{q}_i)$  are characterized by the quantum numbers $n_i$ and $m_i$ and the energy scale $b\equiv\sqrt{P^+\Omega}$. 
The basis is made finite by restricting the number of allowed oscillator quanta in each many-body basis state according to
$
\sum_i\left(2n_i+|m_i|+1\right)\leq N_{\max}.
$
The transverse continuum limit corresponds to $N_{\max}\to\infty$. In addition, we use an ``M-scheme'' basis. That is, our many
body states have well defined values of the total angular momentum projection
$
M_J=\sum_i\left(m_i+s_i\right),
$
where $s=\pm\frac{1}{2}$ is the fermion spin, but they do not have a well-defined total angular momentum $J$.

In BLFQ, we construct our many-body basis in single-particle coordinates. The rationale for doing this is its straightforward
generalization to a basis of many particles. In principle, relative (Jacobi) coordinates could be used, but this process rapidly 
becomes intractable as the particle number is increased, due to the need for proper (anti-)symmetrization of the basis states. 
The disadvantage of using single-particle coordinates is that the center-of-mass (CM) motion of the system
is contained in our solutions. The use of the HO
basis combined with the $N_{\max}$ truncation allows for the exact factorization of the wavefunction into ``intrinsic'' 
and ``CM'' components, even within a truncated basis. 

When the Hamiltonian is expressed in terms of the coordinates $\mathbf{q}\equiv\mathbf{p}/\sqrt{x}$ 
and $\mathbf{s}\equiv\sqrt{x}\mathbf{r}$, exact CM factorization is achieved for all eigenstates,
even in a basis with arbitrary numbers of sectors, which is the reason for the introduction of these coordinates \cite{krakow}.
The spurious CM excited states 
can be removed from the low-lying spectrum by adding a Lagrange multiplier proportional to $H_{CM}$ to the Hamiltonian to get
$
H'=H+\lambda\left(H_{CM}-2b^2I\right),
$
where $H\equiv H_{LC}$. 
In practice, one selects $\lambda$ to be large enough that $2\lambda b^2$ is well above the excitation spectrum of interest. 
Demonstrations of the exact CM factorization within BLFQ are given in Refs. \cite{krakow,yang}.

\section{Two-Body Effective Interaction}
\label{sec:effective}
We truncate the Fock space to include only $\Ket{e^+e^-}$ and $\Ket{e^+e^-\gamma}$ states. We wish to 
formulate an effective potential acting only in the $\Ket{e^+e^-}$ space that includes the effects 
generated by the $\Ket{e^+e^-\gamma}$ space.
The total Hamiltonian can be expressed as
$
\Bra{f}H_{\tot}\Ket{i}=\Bra{f}\left(H_0+H_{\inst}+H_{\eff}\right)\Ket{i},
$
where states $\ket{i}$ and $\ket{f}$ are states in the $\Ket{e^+e^-}$ space.
The basis states $\ket{i}$ and $\ket{f}$ are eigenstates of the free Hamiltonian (i.e. $H_0\ket{i}=\epsilon_i\ket{i}$) with eigenvalue
$
\epsilon_i=\sum_j(\mathbf{p}_j^2+m_j^2)/x_j,
$
where the sum runs over particles (of mass $m_j$) in the state $\ket{i}$.
The instantaneous photon exchange term, the two-body interaction $H_{\inst}$, is the only term in the light-cone Hamiltonian directly connecting two states 
within the $\Ket{e^+e^-}$ sector.
We choose the Bloch form of the effective Hamiltonian.
The Bloch Hamiltonian \cite{dipankar} is given by:
\begin{align}
\Bra{f}H_{\eff}\Ket{i}=\frac{1}{2}\sum_{n}\Bra{f}H\Ket{n}\Bra{n}H\Ket{i}\left[\frac{1}{\epsilon_i-\epsilon_n}+
\frac{1}{\epsilon_f-\epsilon_n}\right].
\label{eq:bloch}
\end{align}
Since we are interested in primarily the effects of repeated photon exchange, we will only include those combinations of 
terms in Eq. (\ref{eq:bloch}) which generate the photon exchange. We neglect the combinations 
which result in the photon being emitted and absorbed by the same fermion. That is, we do not incorporate the fermion self-energy, 
and therefore no fermion mass renormalization is necessary in this model. In addition,
we work with unit-normalized eigenstates and a fixed value of the coupling constant.

One can show that $H_{\inst}$ is completely cancelled by a corresponding term in $H_{\eff}$ leaving ($\alpha=g^2/4\pi$)
\begin{align}
H_{\inst}+H_{\eff}&=\frac{\alpha}{K}\sum_{\bar{\alpha}_1\bar{\alpha}_1'\bar{\alpha}_2\bar{\alpha}_2'}\delta^{j_1'+j_2'}_{j_1+j_2}
b_{\bar{\alpha}_1'}^\dagger d^\dagger_{\bar{\alpha}_2'}d_{\bar{\alpha}_2}b_{\bar{\alpha}_1}\sqrt{x_1x_2x_1'x_2'}\nonumber\\
&\cdot\int \frac{d^2\mathbf{q}_1}{(2\pi)^2} \frac{d^2\mathbf{q}_1'}{(2\pi)^2} \frac{d^2\mathbf{q}_2}{(2\pi)^2} \frac{d^2\mathbf{q}_2'}{(2\pi)^2}
\frac{(2\pi)^2\delta^{(2)}(\sqrt{x_1}\mathbf{q}_1+\sqrt{x_2}\mathbf{q}_2-\sqrt{x_1'}\mathbf{q}_1'-\sqrt{x_2'}\mathbf{q}_2')}
{(x_1-x_1')\frac{1}{2}\left[\left(\epsilon_i-\epsilon_n\right)+\left(\epsilon_f-\epsilon_n\right)\right]}\nonumber\\
&\cdot\Psi_{n_1}^{m_1}(\mathbf{q}_1)\Psi_{n_2}^{m_2}(\mathbf{q}_2)\Psi_{n_1'}^{m_1'\ast}(\mathbf{q}_1')\Psi_{n_2'}^{m_2'\ast}(\mathbf{q}_2')
S_{\alpha_1,\alpha_2,\alpha_1',\alpha_2'}(\sqrt{x_1}\mathbf{q}_1,\sqrt{x_2}\mathbf{q}_2,\sqrt{x_1'}\mathbf{q}_1',\sqrt{x_2'}\mathbf{q}_2'),
\label{eq:heff}
\end{align}
where we use $\bar{\alpha}_i$ to represent the set of discrete quantum numbers $(j_i,s_i,n_i,m_i)$ and 
$\alpha_i$ to represent the subset $(j_i,s_i)$. The explicit expression for the spinor part, 
$S_{\alpha_1,\alpha_2,\alpha_1',\alpha_2'}(\sqrt{x_1}\mathbf{q}_1,\sqrt{x_2}\mathbf{q}_2,\sqrt{x_1'}\mathbf{q}_1',\sqrt{x_2'}\mathbf{q}_2')$,
as well as details of the evaluation of the highly oscillatory 8D integration and the derivation of (\ref{eq:heff}), are presented in Ref. 
\cite{forthcoming}.

In the energy denominator, we introduce a fictitious photon mass $\mu$ to regulate the expected Coulomb singularity that, while integrable,
introduces numerical difficulties. We then need to examine the physical limit $\mu\to 0$.

Previous studies investigating the problem of positronium on the light-front with a one-photon exchange kernel have noted a small but noticeable
dependence on the ultraviolet cutoff of the theory \cite{pauliwolz,trittmann,muonium,karmanov}. Both Refs. \cite{pauliwolz,karmanov}
state that the origin of this instability can be traced to a 
particular term in the effective interaction (or one-photon exchange kernel) which tends to a non-zero constant (in momentum space) at
asymptotically large momentum transfer, corresponding to a Dirac delta potential in coordinate space. Since the 2D Dirac delta potential well
has no bound states of finite binding energy \cite{deltawell}, this leads to a divergence.
In our calculation, this would result in a ground state energy that does not converge with respect to $N_{\max}$.
If this term is simply dropped, the results become convergent. This procedure has some justification, 
as this divergent piece of the effective interaction will be cancelled when higher Fock
sectors are included \cite{pauliwolz}. 
Below, we consider the continuum limit using only this modified interaction, where convergence can be expected.

\section{Numerical Results}
\label{sec:results}

\begin{figure}[tb]
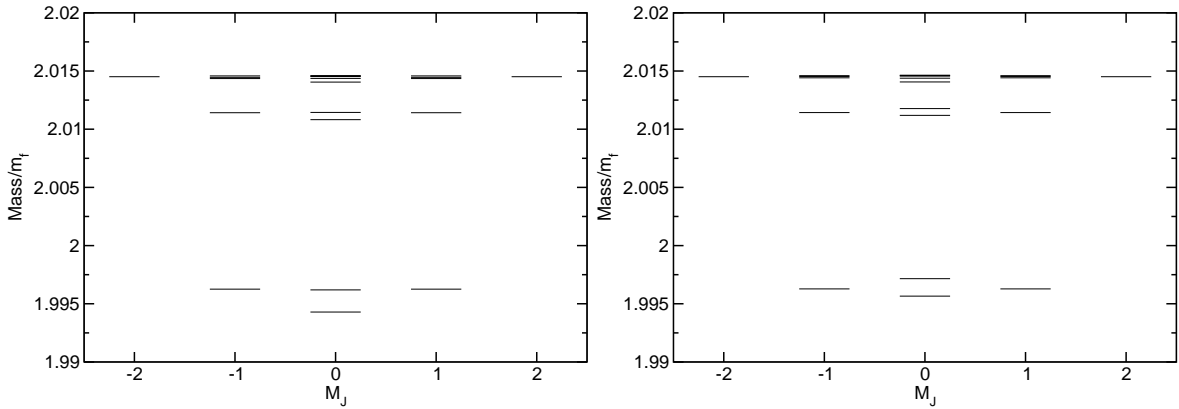

\centering
\includegraphics[width=0.49\textwidth]{spectrum.eps}
\includegraphics[width=0.49\textwidth]{spectrum_mod.eps}
\caption{Representative spectrum of positronium ($\alpha=0.3$) calculated in BLFQ at $K=N_{\max}=19$ and $b=\mu=0.1m_f$,
using the unmodified (left) and modified (right) effective interactions.
The exact energies shown should not be interpreted as final converged results.
Using the unmodified interaction, the approximate
rotational invariance allows for the clear identification of the $1 ^1S_0$, $1 ^3S_1$, $2 ^1S_0$, $2 ^3S_1$,
$2 ^1P_1$, $2 ^3P_0$, $2 ^3P_1$ and $2 ^3P_2$ states of the positronium system (see text for details). Using the modified interaction
(right), the approximate rotational invariance is more strongly broken.}
\label{fig:spectrum}
\end{figure}

We now show the results of our numerical diagonalization of the LFQED Hamiltonian using the effective two-body interaction described
above. Our numerical results were obtained using the Hopper Cray XE6 and Edison Cray XC30 at NERSC. 
In principle the basis energy scale $b$ is arbitrary, as any complete basis can represent the positronium wavefunction. However, the convergence
rate with respect to $N_{\max}$ does depend strongly on $b$. To find the optimal value of $b$ for the ground state, we plotted the ground state
energy as a function of $b$ at $K=N_{\max}=25$, effectively using $b$ as a variational parameter, and found a minimum at $b=0.4m_f$. This value
of $b$ is thus the optimal one for the convergence of the ground state energy.

A representative spectrum of the unmodified interaction is shown in the left panel of Fig. \ref{fig:spectrum}. These results are produced with
$K=N_{\max}=19$ and $b=\mu=0.1m_f$. The energies shown are only representative and should not be considered
converged or final. The general features of the spectrum shown here are common to any calculation with $K=N_{\max}=19$ and above.
Convergence will be considered below. 

The total angular momentum $J$ of the states can be inferred by examining the multiplet structure of the spectrum 
as it appears in the left panel of Fig. \ref{fig:spectrum}.
The ground state, for example, appears only in the $M_J=0$ calculation, suggesting that it has $J=0$.
We also see a triplet of states above the ground state with $M_J=-1,0,1$, suggesting that these states form a $J=1$ multiplet.
The lack of manifest rotational invariance due to the truncation is seen only in the lack of exact degeneracy between the states in 
this multiplet. The difference, however, is quite small, being approximately $1\%$ of the binding energy.
We therefore identify the low-lying $J=0$ and $J=1$ multiplets in
the BLFQ spectrum as the expected $1 ^1S_0$ and $1 ^3S_1$ states of positronium.
The remaining  states can be identified, via similar reasoning, to be 
the $2 ^1S_0$, $2 ^3S_1$, $2 ^1P_1$, $2 ^3P_0$, $2 ^3P_1$ and $2 ^3P_2$ states.

In the right panel of Fig. \ref{fig:spectrum}, the spectrum of the modified interaction is shown.
The only difference is that the rotational invariance is more severely broken. Compared to 
the basic interaction, the $M_J=0$ states are shifted upwards, while the $M_J=\pm1$ states remain essentially unchanged.
This results from the {\it ad hoc} dropping of a divergent term in the interaction, as discussed in Sec. \ref{sec:effective}. 
Despite the lack of (near) rotational symmetry in the modified interaction, we will continue to use the same
state identifications used in the unmodified case. 

\begin{figure}[tb]
\centering
\includegraphics[width=0.44\textwidth]{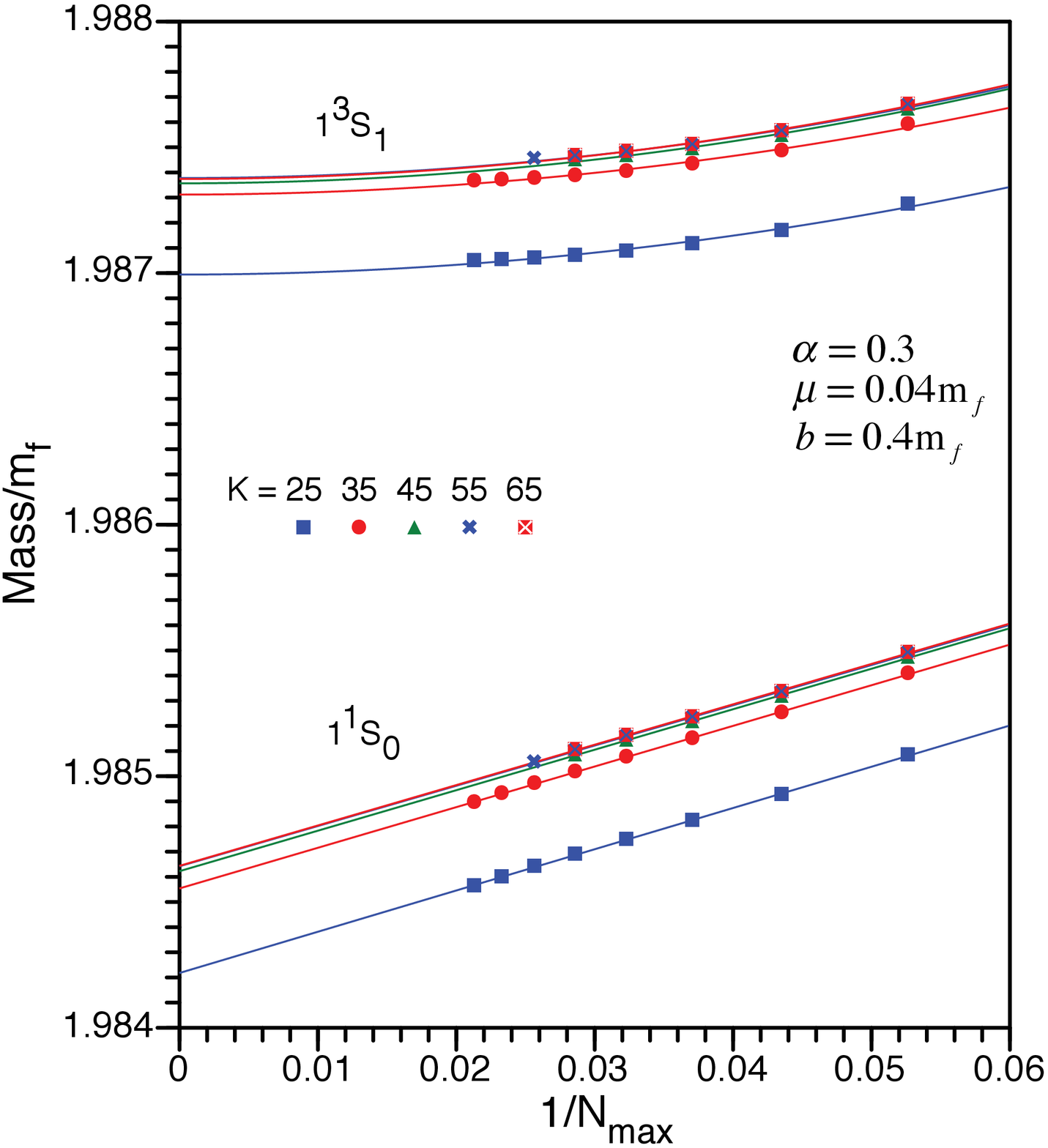}
\includegraphics[width=0.42\textwidth]{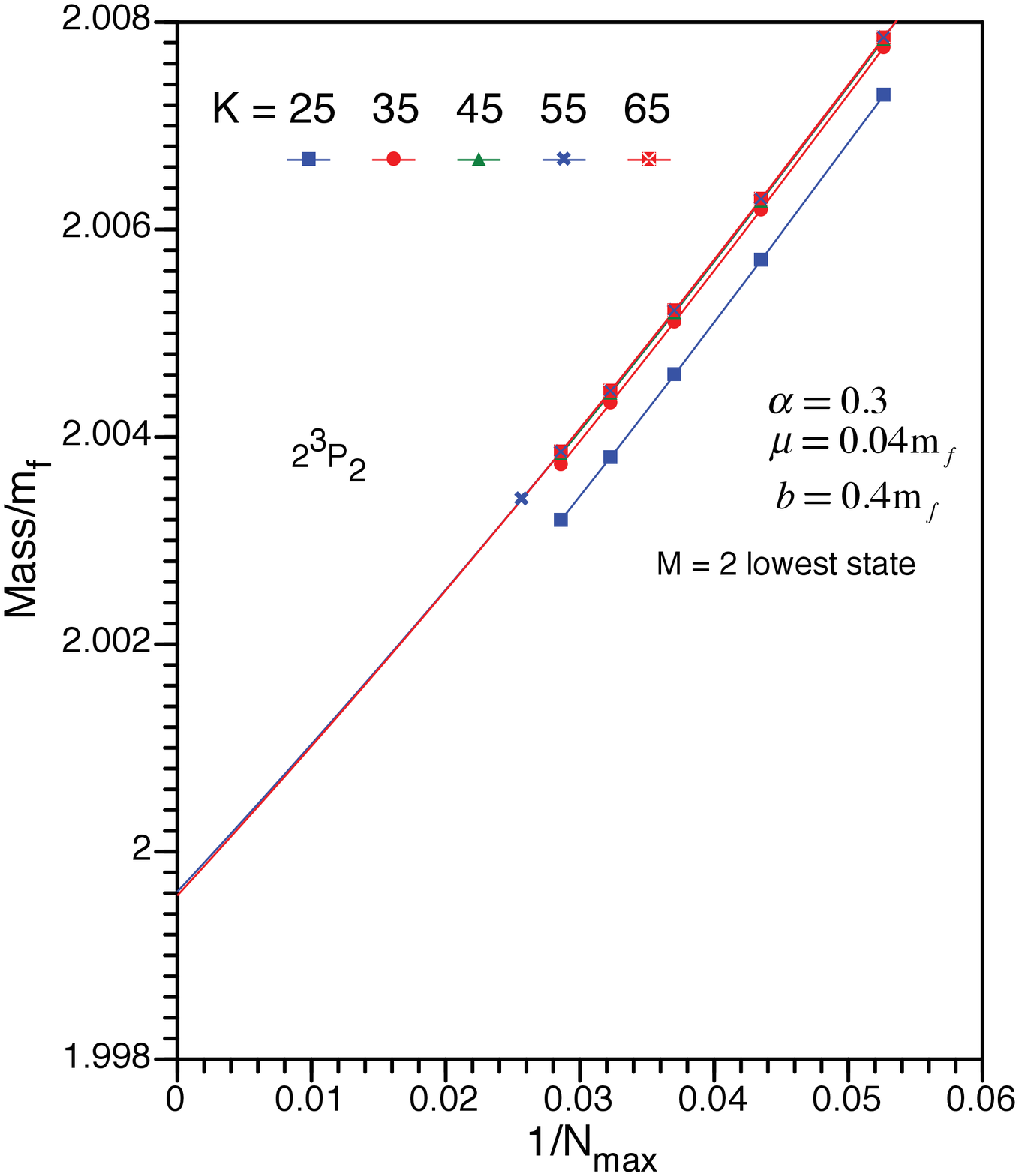}

\caption{Left panel: Dependence of the $1 ^1S_0$ and $1 ^3S_1$ states on the regulators $N_{\max}$ and $K$ for $\mu=0.04m_f$. Results are plotted as a function of
$1/N_{\max}$ for the indicated values of $K$. Convergence with $K$ is rapid. The continuum limit of $N_{\max}\to\infty$ corresponds to 
$1/N_{\max}\to0$. To extrapolate to the continuum limit, the curves are fit to second order polynomials.
Right panel: Same for $2 ^3P_2$ state.}
\label{fig:ground+hyperfine}
\end{figure}

We now examine the convergence of these states as a function of our regulators $N_{\max}$ and $K$.
In the left panel of Fig. \ref{fig:ground+hyperfine}, we show the dependence of the singlet and triplet ground states on the regulators $N_{\max}$ and $K$ for 
a fixed value of the fictitious photon mass $\mu=0.04m_f$. The results are plotted as a function of $1/N_{\max}$ for various fixed values of $K$
as indicated in the figure. The energies converge rapidly with $K$, as one can see from the fact that the curves for $K=45,55,65$ are nearly overlapping
on this scale. We plot the results as a function of $1/N_{\max}$ so that the continuum
limit ($N_{\max}\to\infty$) corresponds to $1/N_{\max}=0$. To extrapolate to the continuum limit, we fit the curves to simple second order
polynomials, although the singlet ground state has a near linear dependence on $1/N_{\max}$. The energies are seen to change by only $\sim 0.001m_f$
over a large range of from $N_{\max}=19$ to $N_{\max}=50$, indicating convergence. 

For a representative higher-excited state, we consider the $2 ^3P_2$ state, found in our calculation as the lowest state of the $M_J=2$ sector. 
In the right panel of Fig. \ref{fig:ground+hyperfine}, 
we show this state's dependence on our regulators $N_{\max}$ and $K$, for the same fixed value of $\mu=0.04m_f$.
This state also shows rapid convergence with respect to $K$, however the dependence on $1/N_{\max}$ is much steeper than for the ground state,
indicating slower convergence for this state. The reason is that the chosen value of the basis energy scale parameter $b$, which was selected to 
be optimal for the ground state, is sub-optimal for this excited state. In fact, a variational calculation suggests that $b\sim0.1m_f$ would be optimal
for this state. 

\begin{figure}[tb]
\centering
  \includegraphics[width=0.45\textwidth]{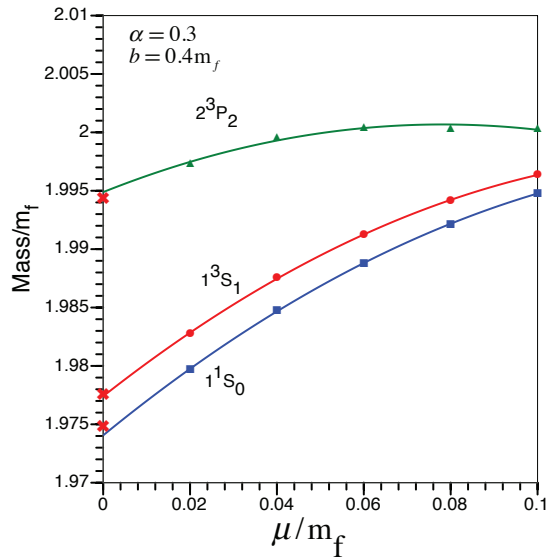}
\caption{The results of the extrapolations in $1/N_{\max}$ (as in Fig. \ref{fig:ground+hyperfine}
for $\mu=0.04m_f$) at a series of values of $\mu$. The physical limit corresponds to $\mu\to0$. The predictions of 
non-relativistic quantum mechanics are indicated as crosses on the vertical axis.}
\label{fig:muto0}       
\end{figure}

To compare our results to the expected Bohr spectrum of positronium, we examine the physical limit of $\mu\to0$. 
In Fig. \ref{fig:muto0}, we plot the results of the $1/N_{\max}$ extrapolations (as in Fig. \ref{fig:ground+hyperfine} 
for $\mu=0.04m_f$) at a series of values of $\mu$.
For comparison, the predicted energies from non-relativistic quantum mechanics are indicated as crosses along the $\mu=0$ axis.
These predictions include the usual first order ``fine structure'' corrections to the Bohr spectrum of positronium. However, we neglect
the correction associated with the electron and positron fluctuating into a virtual photon, since we do not include the $\ket{\gamma}$
sector in our basis.  
To examine the limit $\mu\to0$, we fit the resulting curves to second order polynomials and extrapolate the results to $\mu=0$.
We find excellent agreement with the expected results. It is also interesting to compare our results to the 
Discretized Light-Cone Quantization (DLCQ) results of Ref. \cite{pauliwolz}, also using $\alpha=0.3$. While their method of calculation is different, 
their interaction incorporates all the same physics and assumptions as ours. In the relevant limit, they obtain a ground state
mass of $1.97376m_f$ and a $^3S_1$ state mass of $1.97708m_f$, nearly identical to our results.
Note especially the slight overbinding of the ground state in both methods.

\section{Conclusions and Outlook}
\label{sec:conclusions}
In this work, we have extended our previous results for the mass spectrum of positronium.
We have truncated the Fock space to include only the $\Ket{e^+e^-}$ and $\Ket{e^+e^-\gamma}$
sectors. We further restricted the basis to the $\Ket{e^+e^-}$ sector only by developing a two-body effective interaction,
incorporating the photon exchange effects generated by the $\Ket{e^+e^-\gamma}$ sector.
However, we neglected fermion-self 
energy effects arising from the $\Ket{e^+e^-\gamma}$ sector.
After dropping a divergent term in the two-body effective interaction, we obtained converged results in the continuum limit for the 
$1 ^1S_0$, $1 ^3S_1$ and $2 ^3P_2$ states of positronium. These results agree well with the predictions of 
non-relativistic quantum mechanics.

A straightforward extension of this work is to include a confining interaction between the fermions. The model should
then be applicable to heavy quarkonia systems. A natural choice for the confining potential in BLFQ is quadratic confinement.
Such a quadratic potential is motivated by the phenomenological success of the ``soft wall'' AdS/QCD model \cite{brodsky_prl,brodsky_ads}. 
The effective interaction implemented here would then 
be interpreted as providing leading-order QCD corrections to the semiclassical approximation provided by the AdS/QCD model.
The BLFQ basis is ideally suited to describe the wavefunctions of systems subject to such a confining potential, and we expect convergence with
$N_{\max}$ to be much more rapid than in the positronium calculation presented here.

\begin{acknowledgements}
The authors wish to thank S.~J.~Brodsky, H.~Honkanen, D.~Chakrabarti and V.~A.~Karmanov for fruitful discussions.
This work was supported in part by the Department of Energy under Grant Nos. DE-FG02-87ER40371
and DESC0008485 (SciDAC-3/NUCLEI) and by the National Science Foundation under Grant No
PHY-0904782.
Computational resources were provided by the National Energy Research Supercomputer Center (NERSC),
which is supported by the Office of Science of the U.S. Department of Energy under Contract
No. DE-AC02-05CH11231.

\end{acknowledgements}

\end{document}